\documentclass[aps,prl,twocolumn,twoside,superscriptaddress,nofootinbib,preprintnumbers,floatfix,longbibliography]{revtex4-2}

\makeatletter
\renewcommand\onecolumngrid{%
  \do@columngrid{one}{\@ne}%
  \def\set@footnotewidth{\onecolumngrid}%
}
\makeatother

\usepackage{hyperref}
\usepackage{graphicx}
\usepackage{amsmath}
\usepackage{mathrsfs}
\usepackage{xspace}
\usepackage{amsfonts}
\usepackage{color}   
\usepackage{booktabs,siunitx}

\newcommand{\eq}[1]{Eq.~\eqref{eq:#1}}

\newcommand{\fig}[1]{Fig.~\ref{fig:#1}}

\newcommand{\nn}{\nonumber}

\allowdisplaybreaks[2]

\newcommand{\Ecal}{\mathcal{E}}

\begin{document}
\title{Precision Jet Substructure of Boosted Boson Decays with Energy Correlators}

\author{Anjie Gao}%
\email{anjiegao@stanford.edu}%
\affiliation{Center for Theoretical Physics -- a Leinweber Institute, Massachusetts Institute of Technology, Cambridge, MA 02139, USA}
\affiliation{SLAC National Accelerator Laboratory, Stanford University, Stanford, CA 94039, USA}
\author{Kyle Lee}%
\email{kylel@mit.edu}%
\affiliation{Center for Theoretical Physics -- a Leinweber Institute, Massachusetts Institute of Technology, Cambridge, MA 02139, USA}
\affiliation{Department of Physics, Yale University, New Haven, CT 06511}
\author{Xiaoyuan Zhang}%
\email{xyz2@mit.edu}%
\affiliation{Center for Theoretical Physics -- a Leinweber Institute, Massachusetts Institute of Technology, Cambridge, MA 02139, USA}
\affiliation{Department of Physics, Harvard University, Cambridge, MA 02138, USA}

\begin{abstract}
We initiate the precision study of boosted jet substructure using energy correlators, applying this framework to hadronic Higgs decays. We demonstrate that the two-body decay of the Higgs manifests as a distinct angular peak at $\theta \sim \arccos(1-2/\gamma^2)$ for Lorentz boost factor $\gamma$. We show that infrared scales, such as the dead-cone effect and confinement transition, are also resolved within the boosted distribution. Precision analytic studies of boosted jet substructure may enable precision electroweak studies and open new avenues for new physics searches. 
\end{abstract}

\preprint{MIT-CTP 5996}
\preprint{SLAC-PUB-260120}

\maketitle

\begin{figure*}
\begin{center}
\includegraphics[scale=0.28]{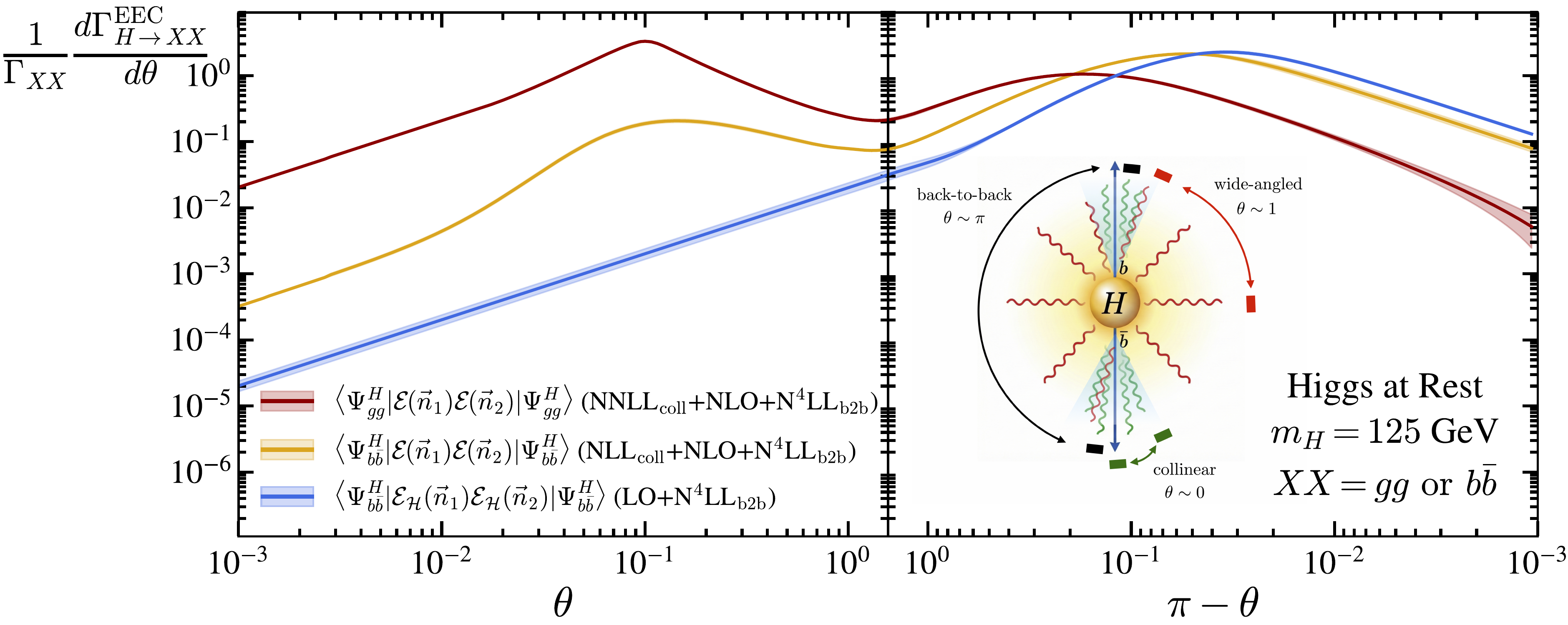}
\end{center}
\vspace{-0.5cm}
\caption{The angular distribution of the energy-energy correlators for hadronic Higgs decays at rest for $gg$ and $b\bar{b}$ decay channels. The angular variable $\theta$ scans across distinct physical regimes: the wide-angle region governed by hard fixed-order QCD ($\theta \sim 1$); the collinear limit ($\theta \ll 1$), characterized by classical $1/\theta$ scaling, DGLAP resummation, and modifications from infrared scales $m_b$ and $\Lambda_{\rm QCD}$; and the back-to-back region ($\theta \to \pi$) dominated by Sudakov double logarithms. The exclusive measurement of $B$-hadron correlations in the $H \to b\bar{b}$ channel is dominated by the back-to-back region.}
\label{fig:full_dist}
\end{figure*}

\emph{\textbf{Introduction.}} Highly energetic partons produced at high-energy colliders undergo successive collinear splittings, generating collimated sprays of particles known as jets. These objects encapsulate the universal behavior of Quantum Chromodynamics (QCD) in the collinear limit. At the energy scales of the Large Hadron Collider and beyond, electroweak bosons ($W, Z, H$) and potential new heavy resonances are frequently produced with extremely boosted kinematics. In this regime, the collimation of their decay products is so significant that these heavy states manifest in detectors as jet-like radiation patterns.

This convergence in appearance necessitates the use of advanced jet substructure techniques to disentangle their origins. This program to analyze radiation patterns within boosted objects was popularized by a seminal study of boosted Higgs decays, reconstructed as jets~\cite{Butterworth:2008iy}. Since its inception, jet substructure program has provided some of the most stringent bounds on new heavy resonances~\cite{CMS:2014hvu,ATLAS:2015xom,CMS:2019qem} and has become a fertile ground for the development of machine learning applications in collider data analysis~\cite{deOliveira:2015xxd,Larkoski:2017jix,Baldi:2016fql,Komiske:2018cqr,Qu:2019gqs,ML4Jets2026,hepmllivingreview}.

In recent years~\cite{Chen:2020vvp,Lee:2022uwt,Chen:2023zlx,Komiske:2022enw,Craft:2022kdo,Andres:2022ovj,Holguin:2022epo,Devereaux:2023vjz,Lee:2023npz,Barata:2023zqg,Holguin:2023bjf,Andres:2023ymw,Lee:2024icn,Bossi:2024qho,Barata:2024wsu,Budhraja:2024tev,Alipour-fard:2024szj,Holguin:2024tkz,Andres:2024ksi,Barata:2025uxp,Barata:2025fzd,Mi:2025abd,Alipour-fard:2025dvp,CMS:2024mlf,ALICE:2024dfl,ALICE:2025igw,Tamis:2023guc,STAR:2025jut,CMS:2024ovv,CMS:2025jam,CMS:2025ydi}, there has been a renewed effort to study jet substructure through the lens of energy correlators measured on a high-energy jet state $|\Psi\rangle$, such as the two-point energy-energy correlator (EEC)~\cite{Basham:1979gh,Basham:1978zq,Basham:1978bw,Basham:1977iq}:
\begin{align}
\langle \Psi| \mathcal{E}(\vec{n}_1) \mathcal{E}(\vec{n}_2) |\Psi\rangle\,,
\end{align}
measuring the correlations between energy-flow operators~\cite{Sterman:1975xv},
\begin{align}
\mathcal{E}(\vec{n}) = \lim_{r \to \infty} \int_0^\infty dt \,r^2 \,n^i\, T_{0i}(t, r\vec{n})\,,
\end{align}
which characterize the energy flux into a calorimeter cell in the direction $\vec{n}$. The inherent simplicity of energy correlators is pushing the boundaries of precision calculations in QCD~\cite{Dixon:2018qgp,Tulipant:2017ybb,Dixon:2019uzg,Duhr:2022yyp,Luo:2019nig,Gao:2020vyx,Yang:2024gcn,Jaarsma:2025tck}. Furthermore, they offer a remarkably clean interpretation of distinct angular regions given by $\theta = \arccos(\vec{n}_1\cdot \vec{n}_2)$  between two detectors, providing a means to clearly discern the physics of confinement, collinear DGLAP evolution, heavy-quark mass effects, medium modifications, and the Sudakov resummation regime. As such correlation functions of asymptotic detectors are natural observables in generic quantum field theories, they are being actively explored in a general field-theoretic context as well. For a comprehensive review, see~\cite{Moult:2025nhu}.

In this letter, we leverage these recent conceptual and precision developments in energy correlators to revisit the foundational goals of the jet substructure program for studying boosted jets. We investigate the correlations produced by boosted states $|\Psi^H_{XX}\rangle$, specifically analyzing the hadronic decays of boosted Higgs bosons by studying the EEC distribution differential in angle $\theta$ as 
\begin{align}
\label{eq:proddecay}
\frac{1}{\sigma_{\rm tot}}\frac{d\sigma^{\rm EEC}_{pp\to H \to XX}}{d\theta} \approx \sigma^{pp \to H} \times \frac{1}{\Gamma_{XX}}\frac{d\Gamma^{\rm EEC}_{H\to XX}}{d\theta}\,,
\end{align}
where $XX \in \{gg, b\bar{b}\}$ are two of the most frequent hadronic decay channels of the Higgs, and $\sigma_{\rm tot}$ is the total cross-section of the $pp\to H\to XX$ decay. 

A key simplification of the Higgs boson is its narrow width, which allows for the factorization between production and decay as shown in~\eq{proddecay}, rendering non-factorizable contributions negligible compared to systems like a top quark~\cite{Hoang:2025kgh}. Consequently, $\sigma^{pp \rightarrow H}$ serves as a proxy for different production channels generating boosted Higgs kinematics. In the \emph{Supplemental Material}, we present next-to-next-to-leading order (NNLO) predictions for $pp\to H+X$ for different Lorentz boost factors $\gamma = E_H/m_H$ as an example of a possible Higgs production process. 

The decay term, $\frac{1}{\Gamma_{XX}}\frac{d\Gamma^{\rm EEC}_{H\to XX}}{d\theta}$, captures the decay dynamics in the boosted frame. The EEC distribution on decays can be written as
\begin{align}
\frac{1}{\Gamma_{XX}}&\frac{d\Gamma^{\rm EEC}_{H\to XX}}{d\theta}  \equiv \frac{1}{E_H^2}\int d^2 \Omega_{\vec{n}_1} d^2 \Omega_{\vec{n}_2} \\
& \times \delta\left(\theta-\arccos(\vec{n}_1 \cdot \vec{n}_2)\right)\left\langle\mathcal{E}\left(\vec{n}_1\right) \mathcal{E}\left(\vec{n}_2\right)\right\rangle_{XX}\,,\nn
\end{align}
where
\begin{align}
\label{eq:EEdef}
&\hspace{-0.3cm}\left\langle\mathcal{E}\left(\vec{n}_1\right) \mathcal{E}\left(\vec{n}_2\right)\right\rangle_{XX} \\
&\hspace{0.4cm}= \frac{ \int \mathrm{~d}^4 x \, e^{i p_H \cdot x} \langle 0| \mathcal{O}_{XX}^\dagger(x) \mathcal{E}\left(\vec{n}_1\right) \mathcal{E}\left(\vec{n}_2\right) \mathcal{O}_{XX}(0)|0\rangle}{\int \mathrm{~d}^4 x \, e^{i p_H \cdot x} \langle 0| \mathcal{O}_{XX}^\dagger(x) \mathcal{O}_{XX}(0)|0\rangle}\,,\nn
\end{align}
with boosted Higgs momentum $p_H= (E_H,\vec{p}_H)$ such that $p_H^2 = m_H^2$. For the decay channels of interest, $\mathcal{O}_{gg} = G_{\mu\nu}^a G^{a,\mu\nu}$ and $\mathcal{O}_{b\bar{b}} = \bar{\psi}_b \psi_b$.

As derived in the \emph{Supplemental Material}, we can relate the EEC of a boosted Higgs to its rest-frame decay using a boost kernel $\mathcal{K}_\gamma$ as
\begin{align}
\label{eq:boost}
\frac{1}{\Gamma_{XX}}\frac{d\Gamma^{\rm EEC}_{H\to XX}}{d\theta} = \int\! d\theta_{\rm rest}\, \mathcal{K}_\gamma(\theta_{\rm rest},\theta) \frac{1}{\Gamma_{XX}}\frac{d\Gamma^{\rm EEC}_{H\to XX}}{d\theta_{\rm rest}}\,.
\end{align}
Thus, our strategy is to first characterize the Higgs decay distribution at rest with precision then apply the appropriate Lorentz boost to match the boost kinematics.

\emph{\textbf{Higgs Decay at Rest.}} The Higgs decay EEC distributions for both the $gg$ and $b\bar{b}$ channels exhibit distinct behaviors across varying angular regions. The relevant energy scales governing these distributions are the Higgs mass $m_H$, the confinement scale $\Lambda_{\rm QCD}$, the angular scale $m_H\sin(\theta/2)$\footnote{For notational simplicity, we use $\theta$ to denote the rest-frame angle $\theta_{\rm rest}$ in this section.}, and, for the $b\bar{b}$ channel, the bottom quark mass $m_b$. As we scan across $\theta$, the angular scale $\mu_{\rm EEC} \sim m_H\sin(\theta/2)$ probes distinct physical regimes defined by these parameters. \fig{full_dist} displays the full angular distribution of various Higgs decay channels and the distinct physical regimes they probe. We detail the physics governing each of these regions below and describe the calculation precision used in each region. 

\textit{The Wide-Angle Region ($\theta \sim 1$):} In the wide-angle region where $\theta \sim 1$, the angular scale is $\mu_{\rm EEC} \sim m_H$. Here, the leading contribution is governed entirely by the hard scale $m_H$, and thus the distribution is described by fixed-order calculations at such UV scale. For both $H\to b\bar{b}$ and $H\to gg$ decays, we compute these contributions to NLO, or $\mathcal{O}(\alpha_s^2)$ accuracy~\cite{Luo:2019nig,Gao:2020vyx}.

\textit{The Collinear Regime ($\theta \ll 1$):} As the angle decreases into the collinear regime, defined by $m_b, \Lambda_{\rm QCD} \ll \mu_{\rm EEC}\sim m_H \theta/2 \ll m_H$, the EEC admits a factorization into a hard function (describing the production of a collinear fragmenting parton) and a massless jet function (describing correlations arising from collinear splittings)~\cite{Dixon:2019uzg,Lee:2022uwt,Chen:2023zlx,Chen:2023zzh,Lee:2025okn}. The collinear splittings generate a classical $1/\theta$ scaling in the correlations, while DGLAP resummation of large logarithms of $\ln \theta$ further induces anomalous scaling determined by twist-2 spin-3 anomalous dimensions, up to effects from $\beta$ functions. For both the $H\to b\bar{b}$ and $H\to gg$ decay channels, we perform the resummation at the highest available logarithmic accuracy: next-to-next-to-leading logarithm (NNLL) for $H\to gg$ and NLL for $H\to b\bar{b}$, the latter being limited by the current availability of the $H\to b\bar{b}$ hard function to NLO~\cite{Corcella:2004xv,Baradaran:2025khn}.

Deeper in the collinear limit, the angular scale $\mu_{\rm EEC} \sim m_H \theta/2$ eventually becomes sensitive to the infrared scales $m_b$ and $\Lambda_{\rm QCD}$. As we approach the massive regime $m_H \theta /2 \sim m_b$, the mass dependence in the collinear splitting can no longer be neglected. While the evolution between $m_H \theta/2 \sim m_b$ and $m_H$ primarily affects the normalization, the shape is governed by the fixed-order massive jet functions~\cite{Craft:2022kdo,Barata:2025uxp} sensitive to mass-dependent collinear splittings. Crucially, this captures the suppression of radiation around a massive emitter known as the dead-cone effect. Consequently, the singular $1/\theta$ scaling is regulated by mass effects, eventually turning over to a linear reduction $\frac{d\Gamma^{\rm EEC}}{d\theta} \propto {\rm const}\times\theta$ as $\theta/2 \lesssim m_b/m_H$. Unless one explicitly selects $g\to b\bar{b}$ splittings~\cite{Barata:2025uxp} inside gluon jets, such mass effects are visible only for $b$-jets produced in $H\to b\bar{b}$ decay. Such mass-dependence of the jet function is only known to NLO, which we use in the massive regime for $b\bar{b}$ decay. 

Finally, at the deepest infrared scales where $m_H\theta/2 \sim \Lambda_{\rm QCD}$, the invariant mass of the splittings approaches the hadronization scale. Here, the production and correlation of the dihadron state are described by moments of dihadron fragmentation functions~\cite{Lee:2025okn,Chang:2025kgq,Kang:2025zto} and cannot be reduced to perturbative objects.\footnote{Although experimental measurements are always performed on hadronic final states, when $m_H \theta \gg \Lambda_{\rm QCD}$, the parton-to-hadron transition occurs on separated branches. In such cases, energy conservation implies that the sum over hadronic states acts effectively as the identity operator, allowing correlations to be written in terms of partonic degrees of freedom.} While these moments are non-perturbative, their evolution follows a perturbative kernel. When $\theta$ falls below the confinement scale, i.e., $m_H\theta/2 \ll \Lambda_{\rm QCD}$, the correlations saturate to a constant value, reflecting a uniform distribution of hadrons. Due to the Jacobian factor $\theta$, this results in a linear scaling $\frac{d\Gamma^{\rm EEC}}{d\theta} \propto {\rm const} \times \theta$. We model this behavior by freezing the $\theta$ dependence of the distribution, consistent with the asymptotic behavior of the moments of dihadron fragmentation functions~\cite{Lee:2025okn,Chang:2025kgq}. Thus, the collinear region exhibits a dramatic modification of scaling driven by additional scales like $m_b$ and $\Lambda_{\rm QCD}$, effectively resolving QCD scales. On the other hand, within conformal field theory, the classical scaling $1/\theta$ remains to arbitrarily small angles~\cite{Hofman:2008ar,Kologlu:2019mfz,Dempsey:2025yiv} due to a lack of such additional mass scales.

\begin{figure}[!t]
\begin{center}
\includegraphics[width=0.48\textwidth]{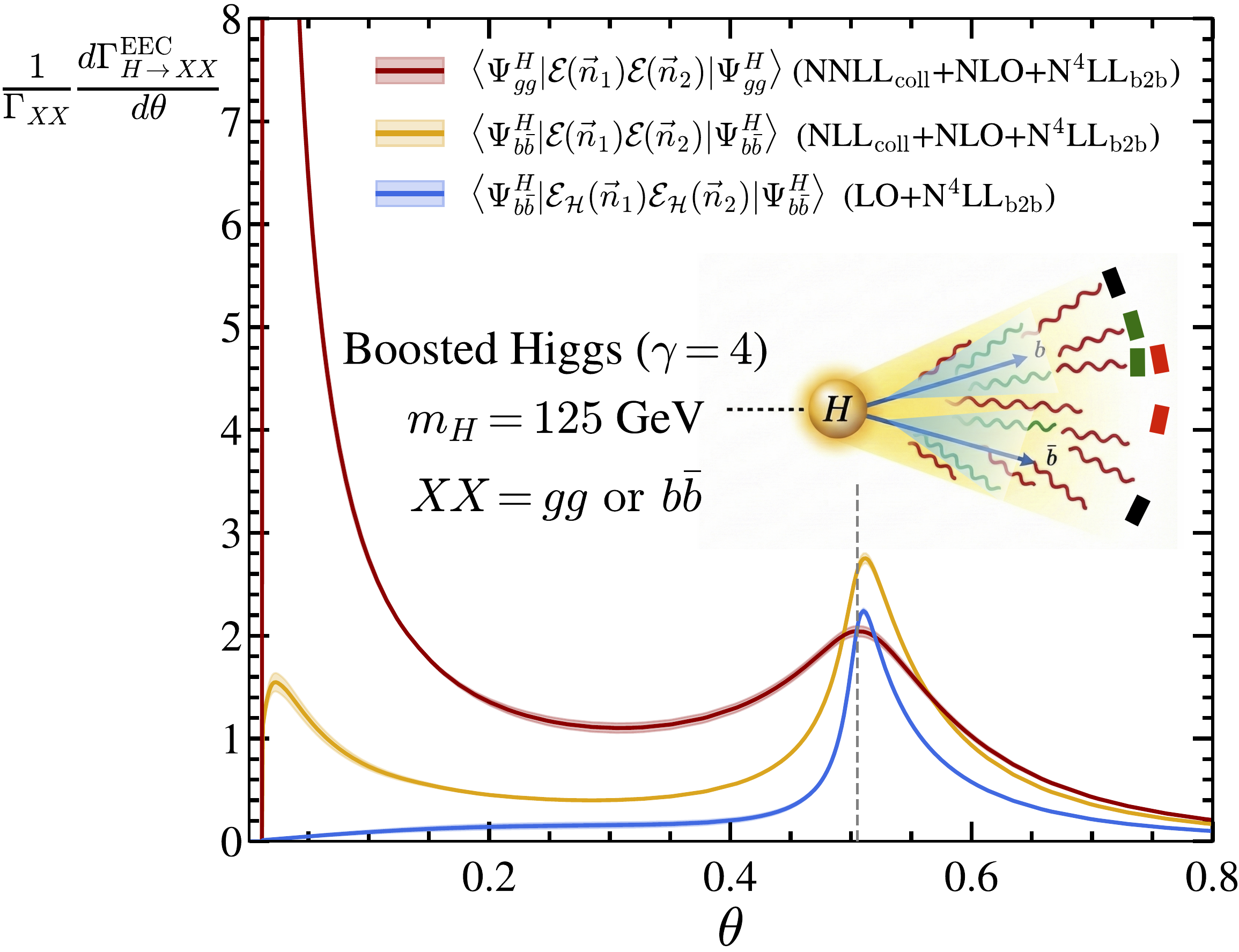}
\end{center}
\vspace{-0.5cm}
\caption{The angular distribution of the energy-energy correlators for a boosted Higgs boson with $m_H = 125$ GeV and $\gamma = 4$ for $gg$ and $b\bar{b}$ decay channels. We compare the inclusive $gg$ and $b\bar{b}$ decay channels, as well as the exclusive measurement of correlations between $B$-hadrons. All distributions exhibit a characteristic peak near $\theta_{\rm peak} \approx \arccos(1-2/\gamma^2)$. The vertical dashed line indicates this kinematic value, corresponding to the boosted topology of the back-to-back two-body decay in the rest frame.}
\label{fig:boost}
\end{figure} 
\textit{The Back-to-Back Regime ($\pi-\theta \ll 1$):} In the back-to-back region, $\theta \to \pi$, the angular scale becomes $\mu_{\rm EEC} \sim m_H (\pi-\theta)/2$, making the distribution sensitive to large logarithms of $\ln(\pi-\theta)$. In contrast to the single logarithmic series of the collinear limit, this region is governed by Sudakov double logarithms arising from both soft and collinear divergences. In the region $m_b, \Lambda_{\rm QCD} \ll m_H (\pi-\theta)/2 \ll m_H$, the Sudakov resummation is required~\cite{Moult:2018jzp,Ebert:2020sfi,Duhr:2022yyp}. Although we carry out N${}^4$LL resummation below, as an illustration, at the LL order, this takes a simple form:
\begin{align}
\label{eq:b2b}
&\frac{2}{\pi-\theta}\frac{1}{\Gamma_{XX}} \frac{d \Gamma_{H\to XX}^{\rm EEC}}{d \theta}\bigg|_{\theta\to \pi} \\
&\simeq H(m_H) \int_0^{\infty} \frac{b d b}{2} J_0(b\,m_H(\pi-\theta)/2) \exp[-S^{XX}_{\mathrm{LL}}(b, m_H)],\nn
\end{align}
where $H(m_H)$ is the hard function and $S^{XX}_{\mathrm{LL}}(b, m_H)=\int_{\mu_b}^{m_H} \frac{d\mu}{2\mu} \Gamma_{\rm cusp}^{XX}(\alpha_s(\mu)) \ln (m_H^2/\mu^2)$ denotes the LL Sudakov exponent, with $\mu_b = 2e^{-\gamma_E}/b$. To leading-logarithmic order, $ \Gamma_{\rm cusp}^{gg}(\alpha_s(\mu)) \approx \alpha_sC_A/\pi \equiv \frac{\alpha_s}{4\pi}\Gamma_{(0)}^{gg}$ and $\Gamma_{\rm cusp}^{b\bar{b}}(\alpha_s(\mu)) \approx \alpha_sC_F/\pi \equiv  \frac{\alpha_s}{4\pi}\Gamma_{(0)}^{b\bar{b}}$.

Physically, this regime corresponds to the formation of a color flux tube or string between the recoiling partons~\cite{Alday:2007mf}. This gives rise to the logarithmic growth of anomalous dimension with cusp anomalous dimensions. Using a saddle-point approximation, one can show that the integrand is dominated at LL by an impact parameter $b_*$ where $S_{\rm LL}^{XX}(b_*)$ is maximized. This is given by $b_* = b_0/m_H \exp[4\pi / (\alpha_s(m_H)(\Gamma_{(0)}^{XX} + 2\beta_0))]$.
 When the argument of the Bessel function $J_0$ becomes constant,\footnote{Note that we use the LL expression $\alpha_s(m_H) = 4\pi/(\beta_0 \ln (m_H^2/\Lambda^2))$ in the last line, where $\Lambda$ is the 1-loop Landau pole scale.}
 \begin{align}
 \label{eq:onset}
\hspace{-0.2cm} b_* m_H(\pi-\theta_{\rm onset})/2 &\sim 1\\
m_H(\pi-\theta_{\rm onset})/2 &\sim  \frac{m_H}{2e^{-\gamma_E}} \exp\left[-\frac{4\pi}{ \alpha_s(m_H)(\Gamma_{(0)}^{XX} + 2\beta_0)}\right]\nn\\
&\sim m_H^{\Gamma_0/(\Gamma_0+\beta_0)}\Lambda^{\beta_0/(\Gamma_0+\beta_0)}\,,\nn
 \end{align}
the right-hand side of~\eq{b2b} begins saturating to a constant value. Including the Jacobian factor $2/(\pi-\theta)$ on the left-hand-side, this results in a linear decay of $\frac{1}{\Gamma_{XX}} \frac{d \Gamma_{H\to XX}^{\rm EEC}}{d \theta}$ as $\theta\to \pi$. In particular, note that the onset angle is significantly smaller for $gg$ decay, relative to $b\bar{b}$ decay due to Casimir scaling in $\Gamma_{(0)}^{XX}$.

Unlike the collinear region, where the $1/\theta$ scaling is modified to the linear $\theta$ scaling by the sensitivity to IR scales $m_b$ or $\Lambda_{\rm QCD}$, the Sudakov suppression is sufficiently strong to already induce the linear $\pi-\theta$ scaling. This behavior is therefore also observed in CFTs~\cite{Korchemsky:2019nzm,Kologlu:2019mfz,Dempsey:2025yiv}. Consequently, the impact of additional IR scales is less striking in the back-to-back region. We assume the constant values (the right-hand side of~\eq{b2b}) are set by the perturbative Sudakov suppression and perform resummation to N${}^4$LL accuracy\footnote{For $H\to gg$ channel, this includes contribution from linearly polarized gluons~\cite{Ebert:2020sfi,Zhu:2025ixc}. For gluon jet functions, we use the `improved gluon jet function' scheme in~\cite{Gao:2024wcg}.}, extrapolating this behavior into the deep infrared. 
The estimation of scale uncertainties follow the convention of~\cite{Stewart:2013faa,Gao:2024wcg}.
The transition to this constant behavior occurs significantly earlier than $\Lambda_{\rm QCD}$ scale $m_H (\pi-\theta_{\rm onset})/2 \gg \Lambda_{\rm QCD}$, especially for $H\to gg$ channel as shown in~\eq{onset}. It would still be interesting to study the exact impact of the presence of the $\Lambda_{\rm QCD}$ scale in the back-to-back region, which would describe the regime where the string eventually breaks, screening the color connection between the back-to-back jets and resulting in color-singlet hadrons. On the other hand, we expect $m_b$ scale to be important for $H\to b\bar{b}$ decay channel as $m_H (\pi-\theta_{\rm onset})/2 \lesssim m_b$ there and makes massive jet function in the back-to-back region important near the onset region. Such mass-dependent jet function was computed to NLO recently~\cite{Dai:2023rvd,vonKuk:2024uxe,Aglietti:2024zhg}, but we do not include its effect here.

\textit{Measurement only on $B$ hadron correlations:} Lastly, rather than performing measurements inclusively on all hadronic decays, we may choose to measure correlations only between $B$ hadrons.\footnote{The energy-flow operator $\mathcal{E}_{\mathcal{H}}$ is defined to measure the energy of heavy hadrons, where $\mathcal{H}$ represents the set of all hadronic states with a heavy quark $Q$ or anti-quark $\bar{Q}$ quantum number.} This is achieved using the track function formalism~\cite{Li:2021zcf,Chang:2013rca,Lee:2023xzv,Lee:2023tkr} for heavy hadrons~\cite{Barata:2025uxp}, treating $b$-flavor as a quantum number~\cite{Georgi:1990um,Isgur:1990yhj,Isgur:1989vq,Manohar:2000dt}. Due to the sensitivity to the perturbative scale $m_b$, such track functions are perturbatively computable up to $\mathcal{O}(\frac{\Lambda_{\rm QCD}}{m_b})$, and we utilize their two-loop expressions~\cite{Mitov:2004du,Melnikov:2004bm} augmented with three-loop logarithms predicted by its renormalization structure. Due to heavy-quark flavor symmetry, $B$ hadrons originate primarily from  $b$ and $\bar{b}$ fragmentations. As the $b\bar{b}$ decay channel is dominated by $b\bar{b}$ pairs produced back-to-back, we intuitively expect the correlations exclusively between $B$ hadrons to be sensitive primarily to the back-to-back singular region. Indeed, at LO, the collinear region of $B$-hadron-only correlations exhibits no singular behavior. While $g\to b\bar{b}$ splittings can induce collinear enhancement at NLO, their contributions will be suppressed. Therefore, we focus mainly on the back-to-back region calculations, which we perform up to N${}^4$LL accuracy.

Since $B$-hadron correlations in $b\bar{b}$ decay constitute a subset of the inclusive correlations, one might expect them to lie strictly below the inclusive baseline across all angles (especially given both are perturbative predictions with positive energy particles). However, we observe a crossover between the two predictions. This behavior again suggests that $m_b$ effects in the jet function are important and must be accounted for near the scale $m_H(\pi-\theta) \sim m_b$, where the crossover occurs.

\begin{figure}[!t]
\begin{center}
\includegraphics[width=0.48\textwidth]{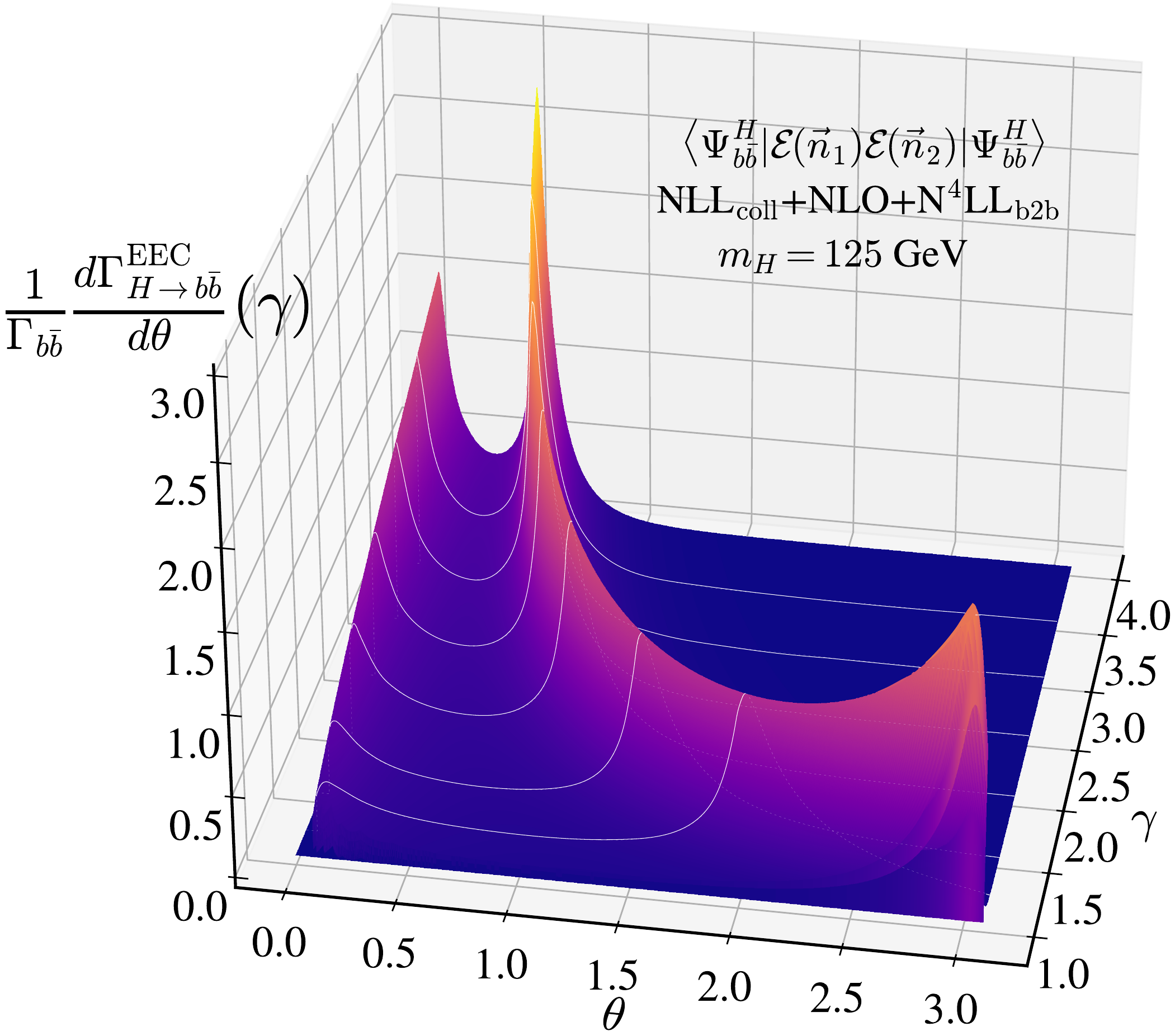}
\end{center}
\vspace{-0.5cm}
\caption{The evolution of the energy-energy correlator distribution for the $H \to b\bar{b}$ decay channel under varying Lorentz boosts $\gamma$. As the boost increases, the characteristic peak shifts to smaller angles, reflecting the increasing collimation of the decay products into a single jet-like topology.}
\label{fig:boost3D}
\end{figure}

\emph{\textbf{Boosted Higgs Decay.}} With precision predictions for the Higgs at rest, we perform the boost using~\eq{boost}. Fig.~\ref{fig:boost} shows the boosted Higgs distribution ($m_H = 125$~GeV) at $\gamma = E_H/m_H = 4$ for different decay channels. As explicitly derived in the \emph{Supplemental Material},  a delta function in the back-to-back region of the rest-frame distribution, $\delta(\pi-\theta)$, transforms under a boost into a peak at $\theta_{\rm peak}= \arccos(1- 2/\gamma^2)$. There, we further argue that this peak structure is expected for the back-to-back resummation EEC distribution with large support near $\theta\sim \pi$ as well. We observe that all curves, namely inclusive $gg$ and $b\bar{b}$ decays as well as the exclusive $b\bar{b}$ measurement, exhibit peaks near this location, though their shapes differ subtly. This peak is a signature of the two-prong structure of the boosted Higgs, reflecting the transformation of a rest-frame, back-to-back two-body decay into a single, collimated jet. At even smaller angles for the inclusive case, we see the collinear peak either from the dead-cone suppression in $H\to b\bar{b}$ or the confinement scale in $H\to gg$, both still preserved under the boost.

In \fig{boost3D}, we isolate the $H\to b\bar{b}$ channel and vary the boost factor $\gamma$. The plot clearly illustrates the shift of peak locations to smaller $\theta$ with increasing $\gamma$, consistent with the increased collimation expected for boosted, jet-like topologies. At hadronic colliders, angular separations are often quantified using the longitudinally boost-invariant distance $\Delta R \equiv \sqrt{\Delta \eta^2 + \Delta \phi^2}$, for which $\Delta R \approx \theta$ holds in the small-angle limit. For highly boosted Higgs jets, most of the distribution is supported at sufficiently small separations that this approximation is applicable. Alternatively, if one tags the precise momentum of the Higgs, one can reconstruct the full angular distribution of its decay even in hadronic colliders.

Crucially, the emergence of this distinct peak at $\theta_{\rm peak} \sim \arccos(1- 2/\gamma^2)$ stands in stark contrast to the behavior of the QCD background. Standard QCD jets, dominated by massless parton branchings, exhibit a characteristic scale-invariant behavior where the EEC distribution scales as $\sim 1/\theta$ (until $\theta$ is small enough to encounter $\Lambda_{\rm QCD}$ or $m_b$ for $b$-jets). The boosted Higgs signal, by contrast, breaks this scale invariance through the mass scale $m_H$, imprinting a `hard' two-prong feature onto the angular distribution.  A precision analytical understanding of such radiation patterns within Higgs jets, facilitated by the EEC as demonstrated here, can assist in the development of high-precision parton showers~\cite{Dasgupta:2020fwr,Forshaw:2020wrq,Giele:2011cb,Hoche:2017iem,Assi:2023rbu,Assi:2025ibi} and improve the performance of machine learning taggers~\cite{Chung:2020ysf,Guo:2020vvt,Khosa:2021cyk,Li:2020grn,Moreno:2019neq,Lin:2018cin} trained on such simulations.

\emph{\textbf{Conclusions and Outlook.}} In this work, we have presented a comprehensive study of the energy-energy correlator for boosted Higgs bosons, integrating precision QCD calculations into the study of boosted jet substructure. Our results demonstrate that the simple transformation properties of energy flow operators under Lorentz boosts render the EEC a particularly robust observable for characterizing boosted resonances, allowing for a straightforward transformation of precision calculations from the rest frame to the boosted distribution.

We showed that the two-prong structure of the Higgs decay, which manifests as a singular back-to-back configuration in the rest frame, maps under the boost to a distinct peak at $\theta \sim \arccos(1-2/\gamma^2)$ inside the jet. Furthermore, we highlighted how fundamental QCD scales, such as the dead-cone suppression in the $H\to b\bar{b}$ channel and the onset of confinement in the $H\to gg$ channel, are imprinted on the angular distribution and preserved in the boosted frame.

Looking forward, this framework opens several avenues for exploration. It will be interesting to investigate whether the precise shape dependence of different decay channels within boosted Higgs jets can be leveraged to statistically tag specific decay modes~\cite{Cavallini:2021vot,ATLAS:2020jwz,Khosa:2021cyk,ATLAS:2018kot,CMS:2013poe,Butterworth:2008sd,Butterworth:2008iy}, potentially enhancing sensitivity to the Higgs Yukawa couplings~\cite{Englert:2015dlp,CMS:2022psv,ATLAS:2024yzu,Michel:2025afc}. As discussed in the End Matter, we can resolve the scalar mass $m_H$ via angular peaks. This suggests potential applications for new physics searches, where the masses of new heavy scalars could be reconstructed directly from the substructure of their decay jets. Furthermore, future lepton colliders~\cite{FCC:2018evy,deBlas:2019rxi,Accettura:2023ked} would provide a pristine environment for the precision measurement of these boosted jet energy correlators, complementing the ongoing program at the Large Hadron Collider.

The theoretical techniques developed here --- combining fixed-order inputs, resummation, and boost kernels --- can be naturally extended to other boosted systems, such as top quarks, hadronically decaying $W/Z$ bosons, or novel resonances predicted by new physics scenarios. Such extensions include considering boosts with polarization information.  
Another important extension is to consider detailed studies for higher-prong decay modes under boost, such as the three-prong structure of top quark decays. Such analyses could enable precision electroweak studies, such as the determination of the top quark mass~\cite{Holguin:2022epo}, or offer novel means to enhance new physics searches, for instance by exploiting interference effects via EEC in polarized decays~\cite{Ricci:2022htc}.

{\it Acknowledgements.}---We thank authors of~\cite{Holguin:HeavyBoson2026} who made us aware of their work as we completed this manuscript and for arranging for coordinated submissions. We thank Philip Harris, Jack Holguin, Sam Homiller, Ian Moult, Francesco Riva, Matthew Schwartz, and Iain Stewart for helpful discussions. 
AG is supported by the United States Department of Energy under contract DE-AC02-76SF00515. KL is supported by funding from the DOE Early Career Award DE-SC0025581. XYZ is supported by the United States Department of Energy under contract DE-SC0013607 and the MIT Pappalardo Fellowship.

\begin{figure*}[!htbp]
\begin{center}
\includegraphics[scale=0.28]{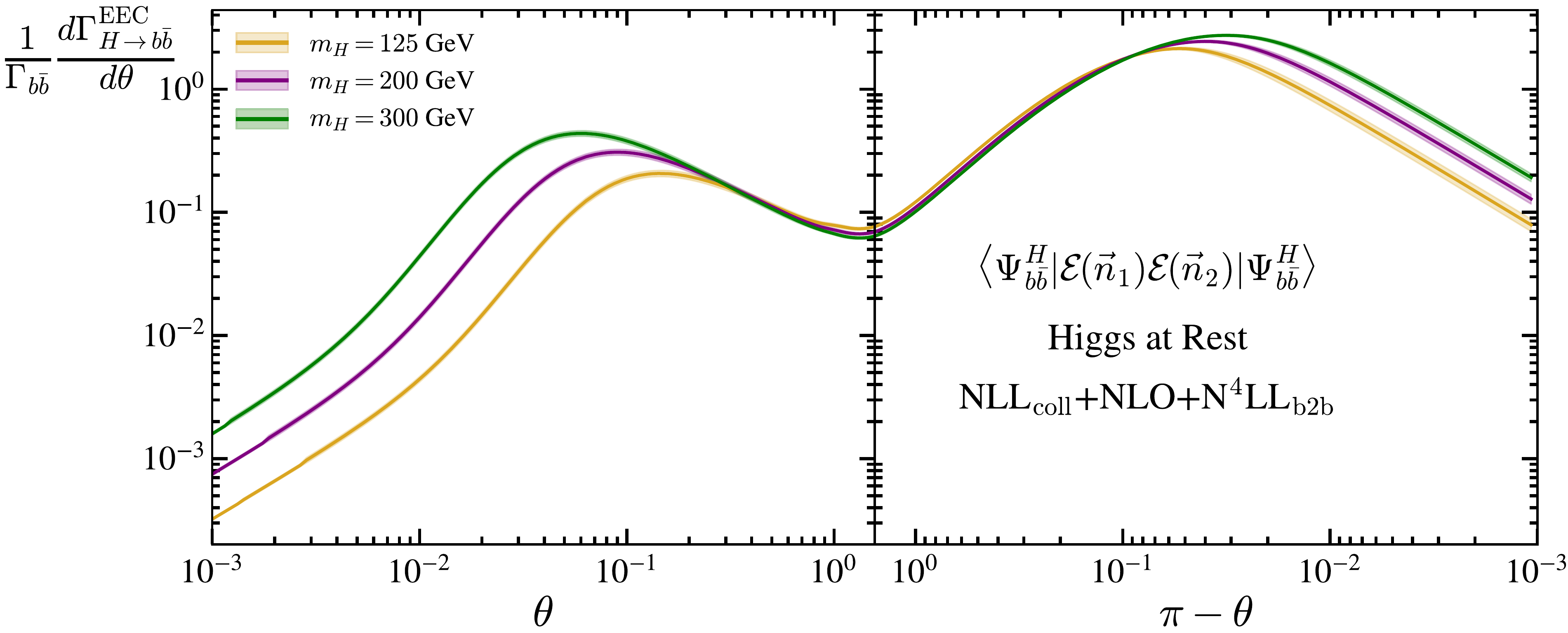}
\end{center}
\vspace{-0.5cm}
\caption{The energy-energy correlator distribution for $H \to b\bar{b}$ decays in the scalar rest frame for varying masses $m_H = 125, 200, 300$ GeV. While the qualitative features of the distribution remain consistent, the angular positions of the peaks shift systematically with the scalar mass. The collinear peak location scales as $m_H \theta/2 \sim m_b$, while the back-to-back peak location is governed by the onset of Sudakov saturation given in~\eq{onset}.}
\label{fig:mHdep}
\end{figure*}

\begin{figure}[!htbp]
\begin{center}
\includegraphics[width=0.48\textwidth]{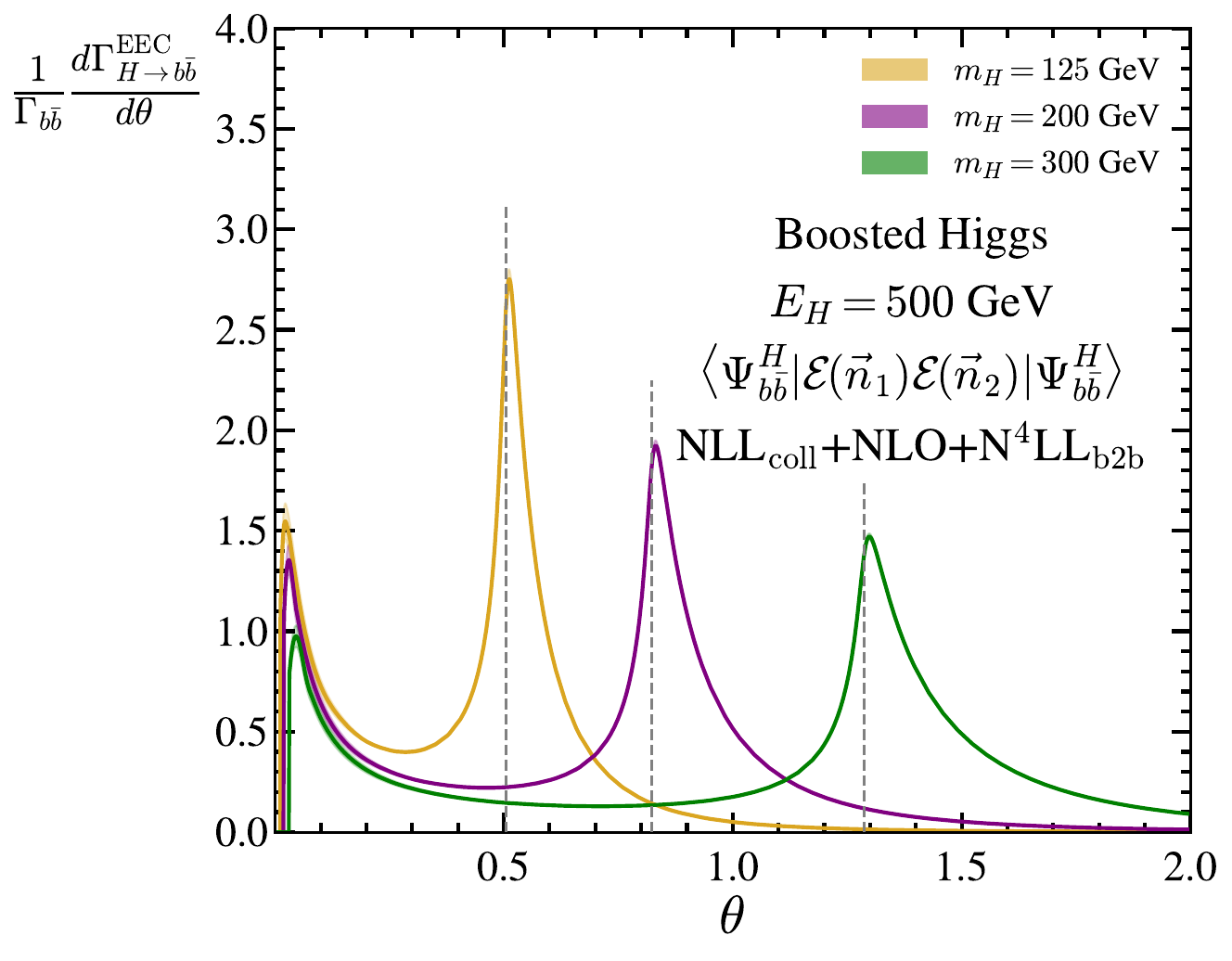}
\end{center}
\vspace{-0.5cm}
\caption{The angular distribution of the EEC for boosted scalars with fixed energy $E_H = 500$ GeV and varying masses $m_H = 125, 200, 300$ GeV. The angular position of the peak depends on the resonance mass. The vertical dashed lines indicate $\arccos(1-2/\gamma^2)$ for different $\gamma = E_H/m_H$.}
\label{fig:boostmHdep}
\end{figure} 
\section*{End Matter}

\emph{\textbf{Scalar Mass Scales Resolved.}} It is also interesting to consider how the hard scale, the mass of the scalar source $m_H$, is imprinted on the EEC distribution. New scalar resonances are motivated in many new physics scenarios, including supersymmetry~\cite{Gunion:1984yn} and two-Higgs-doublet models~\cite{Branco:2011iw}. In \fig{mHdep}, we display the EEC distribution for $H\to b\bar{b}$ decays in the scalar rest frame for varying $m_H = 125, 200, 300$ GeV. While the qualitative features remain consistent, the angular positions of the peaks shift systematically according to the expectations discussed above:
the collinear peak occurs at $m_H \theta/2 \sim m_b$, and the $m_H(\pi-\theta_{\rm onset})/2$ given by~\eq{onset} for the back-to-back region, which exhibits only a mild dependence on $m_H$. Therefore, $m_H$ roughly sets the overall scale for the angle in EEC distribution.

In \fig{boostmHdep}, we examine the imprint of the mass scale $m_H$ in the boosted regime, holding the total energy of the scalar fixed at $E_H= 500$ GeV. As we vary the mass $m_H = 125, 200, 300$ GeV, the corresponding Lorentz boost factor $\gamma = E_H/m_H$ changes, resulting in a distinct shift of the peak location $\theta_{\rm peak} \sim  \arccos(1- 2/\gamma^2)$. Consequently, for a jet of fixed energy, the angular location of this peak serves as a direct proxy for the resonance mass, demonstrating the capability to tag the underlying scalar source.


\bibliography{EEC_ref.bib}{}
\bibliographystyle{apsrev4-1}
\newpage
\onecolumngrid
\section*{Supplemental Material}
\label{sec:supplemental}

\subsection{A. Analytic Derivation of the Boost Kernel and Peak Structure}
In this section, we derive the boost kernel $\mathcal{K}_\gamma$ in \eq{boost}, which maps energy-energy correlator measurements of a hadronic decay from the rest frame of the Higgs to the measurement in its boosted frame. Additionally, we demonstrate that hadronic two-body decays result in a distinct peak at $\theta_{\rm peak} \approx \arccos(1-2/\gamma^2)$ after the boost, where $\gamma$ is the Lorentz boost factor.

\subsubsection*{Boost Kernel Derivation}
To derive the boost kernel, it is convenient to introduce the angular variables 
\begin{align}
  z \equiv \frac{1-\cos\theta}{2}\,,
  \qquad
  z' \equiv \frac{1-\cos\theta'}{2}\,,
\end{align}
where $\theta$ and $\theta'$ represent the opening angles between the calorimeter directions in the rest and boosted frames, respectively. For notational brevity in this section, we denote rest-frame quantities without a prime and boosted-frame quantities with a prime ($'$), such that $\theta = \arccos(\vec{n}_1\cdot \vec{n}_2)$ and $\theta'= \arccos(\vec{n}'_1\cdot \vec{n}'_2)$.

Let $\vec n$ be a unit 3-vector specifying a detector direction in the rest frame, and define the associated null vector
$n^\mu=(1,\vec n)$. When we perform a Lorentz boost with velocity $\vec{\beta}$, characterized by the boost factor $\gamma = 1/\sqrt{1-\beta^2}$ and the Lorentz transformation matrix $\Lambda^\mu_\nu(\vec{\beta})$, this transforms the unit null vector to~
\begin{align}
\label{eq:ntrans}
\Lambda^{\mu}_\nu(\vec{\beta}) n^\nu = \gamma(1+\vec{\beta}\cdot \vec{n}) n'^\mu \equiv \lambda(\vec{\beta},\vec{n}) n'^\mu\,,
\end{align}
where $n'^\mu$ is the resulting unit null vector from the boost and $\lambda(\vec{\beta},\vec{n})$ is the associated rescaling factor. (Henceforth, we will omit the explicit dependence on $\vec{\beta}$ in the argument of $\lambda$ for brevity.) Such rescaling and the transformation of $n^\mu \to n'^\mu$ via the Lorentz transformation $\Lambda$ transforms the energy flow operator $\Ecal(\vec n)$ as~\cite{Belitsky:2013bja,Riembau:2025isw}
\begin{align}
  \label{eq:op_trans}
  U^\dagger(\Lambda)\,\Ecal(\vec n')\,U(\Lambda)
  = \lambda(\vec n)^3\,\Ecal(\vec n)\,.
\end{align}

Therefore, the two-point energy correlators in the boosted frames are related to those of the rest frame as
\begin{align}
  \label{eq:corr_trans}
  \langle \Psi'|\Ecal(\vec n_1')\Ecal(\vec n_2')|\Psi'\rangle
  &=  \langle \Psi|U^\dagger(\Lambda)\Ecal(\vec n_1')\Ecal(\vec n_2')U(\Lambda)|\Psi\rangle\nn\\
  &=  \langle \Psi|\left(U^\dagger(\Lambda)\Ecal(\vec n_1')U(\Lambda)\right) \left(U^\dagger(\Lambda)\Ecal(\vec n_2')U(\Lambda)\right) |\Psi\rangle\nn\\
  &=  \lambda(\vec n_1)^3\lambda(\vec n_2)^3\,\langle \Psi|\Ecal(\vec n_1)\Ecal(\vec n_2)|\Psi\rangle\,.
\end{align}
where our states have definite momentum as
\begin{align}
\langle \Psi|\Ecal(\vec n_1)\Ecal(\vec n_2)|\Psi\rangle&=\frac{ \int \mathrm{~d}^4 x \, e^{i p_H \cdot x} \langle 0| \mathcal{O}^\dagger(x) \mathcal{E}\left(\vec{n}_1\right) \mathcal{E}\left(\vec{n}_2\right) \mathcal{O}(0)|0\rangle}{\int \mathrm{~d}^4 x \, e^{i p_H \cdot x} \langle 0| \mathcal{O}^\dagger(x) \mathcal{O}(0)|0\rangle}\,,\nn\\
\langle \Psi'|\Ecal(\vec n_1')\Ecal(\vec n_2')|\Psi'\rangle&= \frac{ \int \mathrm{~d}^4 x \, e^{i p_H' \cdot x} \langle 0| \mathcal{O}^\dagger(x)U^\dagger(\Lambda) \mathcal{E}\left(\vec{n}'_1\right) \mathcal{E}\left(\vec{n}'_2\right) U(\Lambda)\mathcal{O}(0)|0\rangle}{\int \mathrm{~d}^4 x \, e^{i p_H' \cdot x} \langle 0| \mathcal{O}^\dagger(x) U^\dagger(\Lambda)U(\Lambda)\mathcal{O}(0)|0\rangle}\,,
\end{align}
with boosted decay particle momentum $p_H'= (E_H,\vec{p}_H) = \gamma m_H(1,\vec{\beta})$ and rest-frame momentum $p_H = (m_H,\vec{0})$ such that $p_H^2 =p_H'^2= m_H^2$.
We can then define the EEC distribution differential only in $z$ (or $z'$) as
\begin{align}
\frac{1}{\Gamma}\frac{d\Gamma}{dz}  &\equiv \frac{1}{m_H^2}\int d^2 \Omega_{\vec{n}_1} d^2 \Omega_{\vec{n}_2}\delta\left(z-\frac{n_1\cdot n_2}{2}\right) \langle \Psi|\Ecal(\vec n_1)\Ecal(\vec n_2)|\Psi\rangle \nn\\
&= \frac{1}{m_H^2}\int 4\pi\, 2\pi \sin\theta d\theta\,\delta\left(z-\frac{n_1\cdot n_2}{2}\right) \langle \Psi|\Ecal(\vec n_1)\Ecal(\vec n_2)|\Psi\rangle \,,\nn\\
\frac{1}{\Gamma}\frac{d\Gamma}{dz'}  &\equiv \frac{1}{\gamma^2 m_H^2}\int d^2 \Omega_{\vec{n}'_1} d^2 \Omega_{\vec{n}'_2}\delta\left(z'-\frac{n'_1\cdot n'_2}{2}\right) \langle \Psi'|\Ecal(\vec n'_1)\Ecal(\vec n'_2)|\Psi'\rangle\,,
\label{eq:zdiffdist}
\end{align}
where for the rest-frame distribution, we used the fact that for a scalar decay source, the decay in its rest-frame is isotropic and thus the  correlator $\langle \Psi | \Ecal(\vec n_1) \Ecal(\vec n_2) | \Psi \rangle$ is only a function of $\theta$. Concretely, $2\pi$ can be thought of as coming from the azimuthal angle of $\vec{n}_2$ with respect to $\vec{n}_1$ and $4\pi$ from integrating over the orientation of $\vec{n}_1$ in the celestial sphere, with respect to which the decay is isotropic. Note also that $m_H^2$ in the overall normalization at rest is the total energy component of the decaying particle at rest. This energy component transforms under boost to $E_H^2=\gamma^2 m_H^2$.

Using \eq{ntrans}, \eq{corr_trans}, and the transformation of solid angle measure under boost $d^2\Omega_{\vec{n}_i'}=\lambda(\vec{n}_i)^{-2}d^2\Omega_{\vec{n}_i}$, we can rewrite the $z'$-differential EEC distribution for the boosted frame, with the boost direction defined by $\vec{\beta}$, as
\begin{align}
  \label{eq:preconv}
\frac{1}{\Gamma}\frac{d\Gamma}{dz'} 
  =&
  \frac{1}{\gamma^2 m_H^2}
  \int\!d^2\Omega_{\vec n_1}d^2\Omega_{\vec n_2}\,
  \lambda(\vec n_1)\lambda(\vec n_2)\,
  \delta\!
  \left(z' - \frac{n_1 \cdot n_2}{2\lambda(\vec n_1)\lambda(\vec n_2)}\right)
  \langle \Psi|\Ecal(\vec n_1)\Ecal(\vec n_2)|\Psi\rangle\nn\\
=&
  \frac{1}{\gamma^2 m_H^2}\int dz 
  \int\!d^2\Omega_{\vec n_1}d^2\Omega_{\vec n_2}\,
  \lambda(\vec n_1)\lambda(\vec n_2)\,
  \delta\!
  \left(z - \frac{n_1 \cdot n_2}{2}\right)\delta\!
  \left(z' - \frac{n_1 \cdot n_2}{2\lambda(\vec n_1)\lambda(\vec n_2)}\right)
  \langle \Psi|\Ecal(\vec n_1)\Ecal(\vec n_2)|\Psi\rangle\nn\\
=&
  \frac{1}{\gamma^2 m_H^2}\int dz 
  \int\!\frac{d^2\Omega}{4\pi}\left[\int\!4\pi \, 2\pi \sin\theta d\theta\,
  \delta\!
  \left(z - \frac{n_1 \cdot n_2}{2}\right)
  \langle \Psi|\Ecal(\vec n_1)\Ecal(\vec n_2)|\Psi\rangle\right] 
  \frac{z}{z'}\delta\!\left(z' - \frac{z}{\lambda(\vec n_1)\lambda(\vec n_2)}\right)\nn\\
=&
  \int dz \left(\,\frac{1}{\gamma^2}\int\!\frac{d^2\Omega}{4\pi}
  \frac{z}{z'}\delta\left(z' - \frac{z}{\lambda(\vec n_1)\lambda(\vec n_2)}\right)\right)
  \frac{1}{\Gamma}\frac{d\Gamma}{dz} \equiv \int dz\, \mathcal{K}_\gamma(z,z')
  \frac{1}{\Gamma}\frac{d\Gamma}{dz}\,.
\end{align}
Going from the second to the third line, we have rewritten the solid angle integration of $\vec{n}_1$ and $\vec{n}_2$ into differential measure in angle $\theta = \arccos(\vec{n}_1\cdot \vec{n}_2)$ and solid angle integration measure $d^2\Omega$ describing the orientation of the $\vec{n}_1$ and $\vec{n}_2$ system with respect to the boost direction $\vec{\beta}$. As the terms in the $[\cdots]$ of the third line do not depend on such orientation for decays of a scalar source, we are able to explicitly carry out the integral to identify the rest-frame EEC distribution $\frac{1}{\Gamma}\frac{d\Gamma}{dz}$ and the boost kernel $\mathcal{K}_\gamma(z,z')$ that relates the rest and boosted frame EEC distributions.

To derive the analytic form of the boost kernel $\mathcal{K}_\gamma(z,z')$, we find it convenient to parameterize the rest-frame vectors $\vec{n}_1$ and $\vec{n}_2$ in terms of $\vec{n}_a$ and $\vec{n}_b$ in the $xz$-plane and a rotation matrix $R$ as $\vec{n}_1 = R n_a$ and $\vec{n}_2 = R n_b$. The rotation $R$ preserves the inner product, $\vec{n}_1\cdot \vec{n}_2 = \vec{n}_a\cdot \vec{n}_b = \cos\theta$, and $\vec{n}_{a,b}$ are parameterized as
\begin{align}
    \vec n_a &= \Bigl(\sin\frac{\theta}{2}, 0, \cos\frac{\theta}{2}\Bigr) \,, 
    \qquad
    \vec n_b = \Bigl(-\sin\frac{\theta}{2}, 0, \cos\frac{\theta}{2}\Bigr) \,.
\end{align}
The rotation matrix $R$ is then parameterized by angles $(\vartheta,\varphi)$ that describe the orientation of the seed vector system $\{\vec{n}_a,\vec{n}_b\}$ system with respect to some boost direction $\vec{\beta}$. Then scaling factors $\lambda(\vec{n}_i) = \gamma(1-\vec{\beta}\cdot\vec{n}_i)$ take the form
\begin{align}
    \lambda(\vec n_1) &= \gamma(1 - \vec{\beta} \cdot \vec n_1) =\gamma(1 - \vec{\beta} \cdot (R\vec n_a)) = \gamma - \sqrt{\gamma^2-1} \left( \sin\frac{\theta}{2} \cos\varphi \sin\vartheta + \cos\frac{\theta}{2} \cos\vartheta \right) \,, \nonumber \\
    \lambda(\vec n_2) &= \gamma(1 - \vec{\beta} \cdot \vec n_2) =\gamma(1 - \vec{\beta} \cdot (R\vec n_b)) = \gamma - \sqrt{\gamma^2-1} \left( -\sin\frac{\theta}{2} \cos\varphi \sin\vartheta + \cos\frac{\theta}{2} \cos\vartheta \right) \,.
\end{align}
By performing the angular integration over such rotations $d^2\Omega = \sin\vartheta d\vartheta d\varphi$, the analytic form of the boost kernel can be derived as
\begin{align}
  \label{eq:Kgamma}
  \mathcal{K}_\gamma(z,z')
  &=\frac{1}{\gamma^2}\int\!\frac{d^2\Omega}{4\pi}
  \frac{z}{z'}\delta\left(z' - \frac{z}{\lambda(\vec n_1)\lambda(\vec n_2)}\right) \nn\\
  &=
  \frac{z^{3/2}\;\Theta(m(z,z'))}{2\pi\,\gamma^2\sqrt{\gamma^2-1}\;
    z'^{5/2}\;\sqrt[4]{(1-z)(1-z')}}\;
  \times
  \begin{cases}
    K\bigl(\sqrt{m(z,z')}\bigr)\,, & 0\leq m(z,z')<1\,,\\[0.15cm]
    \dfrac{1}{\sqrt{m(z,z')}}\,K\bigl(1/\sqrt{m(z,z')}\bigr)\,, & m(z,z')>1\,,
  \end{cases}
\end{align}
where $K(k)$ is the complete elliptic integral of the first kind
\begin{align}
K(k)=\int_0^1 \frac{d t}{\sqrt{\left(1-t^2\right)\left(1-k^2\, t^2\right)}}\,,
\end{align}
and the parameter $m(z,z')$ in~\eq{Kgamma} is given by $z,z',$ and $\gamma$ as
\begin{align}
  \label{eq:mdef}
  m(z,z')  =  \frac{ 2\sqrt{1-z}\sqrt{1-z'} + z + z' + z z'(\gamma^2-1) - 2 }{4\sqrt{1-z}\sqrt{1-z'}} \,.
\end{align}

In particular, the boost kernel takes a simple form in the $z\to0$ and $z\to1$ limits as
\begin{align}
\mathcal{K}_\gamma(z=0,z')
&=\frac{4\gamma^2-1}{3\gamma^2}\delta(z')\,, \nn
\\
\mathcal{K}_\gamma(z=1,z')
&=\frac{\Theta(\gamma^2 z'-1)}{2\gamma^2 \sqrt{\gamma^2-1}\, z'^{5/2}\sqrt{\gamma^2 z'-1}}\,.
\label{eq:Kgammaz0z1}
\end{align}

For $0<z<1$, depending on the value of $m(z,z')$, the boost kernel in~\eq{Kgamma} lies in different branches ($0\leq m(z,z')<1$ and $m(z,z')>1$)\footnote{The condition $m(z,z') < 0$ corresponds to a kinematically forbidden configuration. While the algebraic expression for $m(z,z')$ can be negative for some choice of $z$ and $z'$, such pairs cannot be related by a Lorentz boost and thus lie outside the physical support of the kernel.}. We can solve for the range of $z,z'$ that corresponds to different branches of $m(z,z')$. If we fix $z$ at different values, then we find\footnote{Since $m(z,z')$ is symmetric under $z\leftrightarrow z'$, the discussion below can also be applied for $z$ when fixing $z'$.
}
\begin{itemize}
    \item For $0< z\leq1/\gamma^2$, we always find $m(z,z')<1$, while the condition $m(z,z')\geq0$ gives a constraint on the range of $z'$ as $z_-'\leq z'\leq z_+'$, where we define $z_\pm'$ below.  Therefore, in this regime, the boost kernel always lie in the first branch $0\leq m(z,z')<1$.
    \item For $1/\gamma^2 < z < 1$, imposing the condition $m(z,z')\geq0$ constrains the range of $z'$ to $z_-'\leq z'\leq 1$. Imposing the $m(z,z')<1$ or $m(z,z')>1$ constraints further separates the range of $z'$ into $z_-'\leq z'< z_+'$ and $ z_+' < z \leq 1$, respectively. As $z'\to z_+'$, we have $m(z,z')\to 1$ and the boost kernel becomes divergent but the singularity is integrable there.
\end{itemize}
The boundary values $z_\pm'$ are defined as
\begin{align}
  \label{eq:zpm}
  z'_{\pm}
  =
  \frac{z}{\Bigl(\gamma \mp \sqrt{\gamma^2-1}\,\sqrt{1-z}\Bigr)^2}\,,
\end{align}
which lie between $0$ and $1$, i.e. $0\leq z_-' \leq z_+'\leq 1$.

Finally, in \eq{boost} of the main-text\footnote{Note that in \eq{boost}, the rest and boosted-frame angles are denoted as $\theta_{\rm rest}$ and $\theta$, whereas in this section they are denoted as $\theta$ and $\theta'$, respectively.}, the boost kernel was expressed in terms of the angles $\theta$ and $\theta'$ rather than $z$ and $z'$:
\begin{align}
\frac{1}{\Gamma}\frac{d\Gamma}{d\theta'} = \int\! d\theta\, \mathcal{K}_\gamma(\theta,\theta') \frac{1}{\Gamma}\frac{d\Gamma}{d\theta}\,.
\end{align}
The relation between the boost kernels expressed in different variables is simply given by the Jacobian factor as
\begin{align}
\mathcal{K}_\gamma(\theta,\theta') = \frac{dz'}{d\theta'}\mathcal{K}_\gamma(z,z') = \frac{\sin\theta}{2}\mathcal{K}_\gamma\left(z= \frac{1-\cos\theta}{2},z' = \frac{1-\cos\theta'}{2}\right)\,.
\end{align}

\subsubsection*{EEC Sum Rule and Boost}
Energy and momentum conservation imply that the rest-frame EEC distribution for inclusive hadronic decays satisfies the following sum rule
\begin{align}\label{eq:sum_rule_rest}
\int_0^1\! dz\,\frac{1}{\Gamma}\frac{d\Gamma}{dz} = 2\int_0^1\! dz\,z \,\frac{1}{\Gamma}\frac{d\Gamma}{dz} = 1\,.
\end{align}
Our boost procedure should preserve such conservation for the boost distribution as well. Normalizing the boosted distribution by $E_H^2 = \gamma^2 m_H^2$, as shown in~\eq{zdiffdist}, maintains energy conservation exactly, but introduces a $1/\gamma^2$ boost factor for the momentum conservation formula, yielding
\begin{align}
\int_0^1\! dz'\,\frac{1}{\Gamma}\frac{d\Gamma}{dz'} = 2\gamma^2\int_0^1\! dz'\,z' \,\frac{1}{\Gamma}\frac{d\Gamma}{dz'} = 1\,.
\end{align}

We can explicitly show this by
\begin{align}
\int_0^1 dz' \frac{1}{\Gamma}\frac{d\Gamma}{dz'} &= \int_0^1 dz\left[dz' \mathcal{K}_\gamma(z,z')\right] \frac{1}{\Gamma}\frac{d\Gamma}{dz} =  \int_0^1 \! dz \left[-\frac{2(\gamma^2-1)}{3\gamma^2}\bigl(z-\frac12\bigr)+1\right] \frac{1}{\Gamma}\frac{d\Gamma}{dz}=1\,,\nn\\
\int_0^1 dz' z'\,\frac{1}{\Gamma}\frac{d\Gamma}{dz'} &= \int_0^1 dz\left[dz' \,z'\,\mathcal{K}_\gamma(z,z')\right] \frac{1}{\Gamma}\frac{d\Gamma}{dz} =  \int_0^1 \! dz \frac{z}{\gamma^2} \frac{1}{\Gamma}\frac{d\Gamma}{dz}= \frac{1}{2\gamma^2}\,,
\end{align}
where we have first done the $dz'$ integrals with $\mathcal{K}_\gamma$ in \eq{Kgamma}, then used \eq{sum_rule_rest}.

\subsubsection*{Peak Structure from a Two-Body Decay}
In the rest frame, the products of a two-body decay move back-to-back, characterized by an opening angle $\theta$ near $\pi$, or equivalently $z \approx 1$. At the Born level, the rest-frame distribution is expressed as
\begin{align}
\frac{1}{\Gamma}\frac{d\Gamma}{dz} = \frac{1}{2}\left[\delta(z)+\delta(1-z)\right]\,, \qquad \text{Born-level, Back-to-back decay}
\end{align}
where $\delta(z)$ comes from putting the detectors on the same particle. Let us focus on the $\delta(1-z)$, or the back-to-back correlation contribution.

As established above, when $z$ lies in the interval $1/\gamma^2 < z < 1$, the boost kernel $\mathcal{K}_\gamma(z,z')$ in~\eq{Kgamma} has two distinct branches: $z_-'\leq z'<z_+'$ with $0\leq m(z,z')<1$ or $z_+'<z'\leq1$ with $m(z,z')>1$. For the Born-level back-to-back correlation with $\delta(1-z)$, the first branch has no region of support as $z'_\pm \xrightarrow{z \to 1} 1/\gamma^2$. Consequently, the boost kernel collapses into a single branch with $m(z,z') > 1$. 

Therefore, using~\eq{Kgammaz0z1}, the resulting boosted distribution for the $\frac{1}{2}\delta(1-z)$ component (full distribution must include the $\delta(z)/2$ part as well) is
\begin{align}
\frac{1}{2}\mathcal{K}_\gamma(z=1,z')
=\frac{\Theta(\gamma^2 z'-1)}{4\gamma^2 \sqrt{\gamma^2-1}\, z'^{5/2}\sqrt{\gamma^2 z'-1}}\,, 
\end{align}
which monotonically increases as $z'\to 1/\gamma^2$ (or equivalently $\theta' \to  \arccos(1-2/\gamma^2)$). 
Note that the $z'=1/\gamma^2$ singularity is integrable in $z'$, which is necessary for the sum rule preservation (after including $\delta(z)/2$ component).

In the two-body back-to-back decay, Sudakov resummation replaces the singular $\delta(1-z)$ with a broadened distribution. As discussed in the main-text, this distribution rises toward $z \approx 1$ before reaching a plateau. Therefore, while the primary support remains concentrated near $z \sim 1$, it is no longer simply a delta function. The LL resummation expression for such Sudakov resummation was given explicitly on the right-hand side of~\eq{b2b}.

While the exact shape of the boosted peak depends on the details of the Sudakov suppression, its location is roughly determined by the analytic structure of the kernel.
As $z$ approaches (but does not reach) unity, contributions now arise from both branches of the boost kernel, and the boundary values $z'_\pm(z)$ both approach $1/\gamma^2$. This forces the support of the first branch ($z'_- \leq z' < z'_+$) into a narrow window around $1/\gamma^2$. While the support of the second branch ($z'_+ < z' \leq 1$) remains wide, its contribution is dominated by the region near the lower boundary $z'_+$, where the kernel rises to an integrable singularity. Since the rest-frame distribution is concentrated near $z \sim 1$, the convolution is dominated by the kernel's behavior near these boundaries, resulting in a peak in the boosted frame near $z' \sim 1/\gamma^2$. This expectation explains the feature observed in the main text, where the hadronic decay peak appears at $\theta'_{\rm peak} \approx \arccos(1-2/\gamma^2)$.

\subsection{B. Higgs Production via $pp\to H+X$}
\label{app:prod}
\begin{figure}[!t]
\begin{center}
\includegraphics[width=0.5\textwidth]{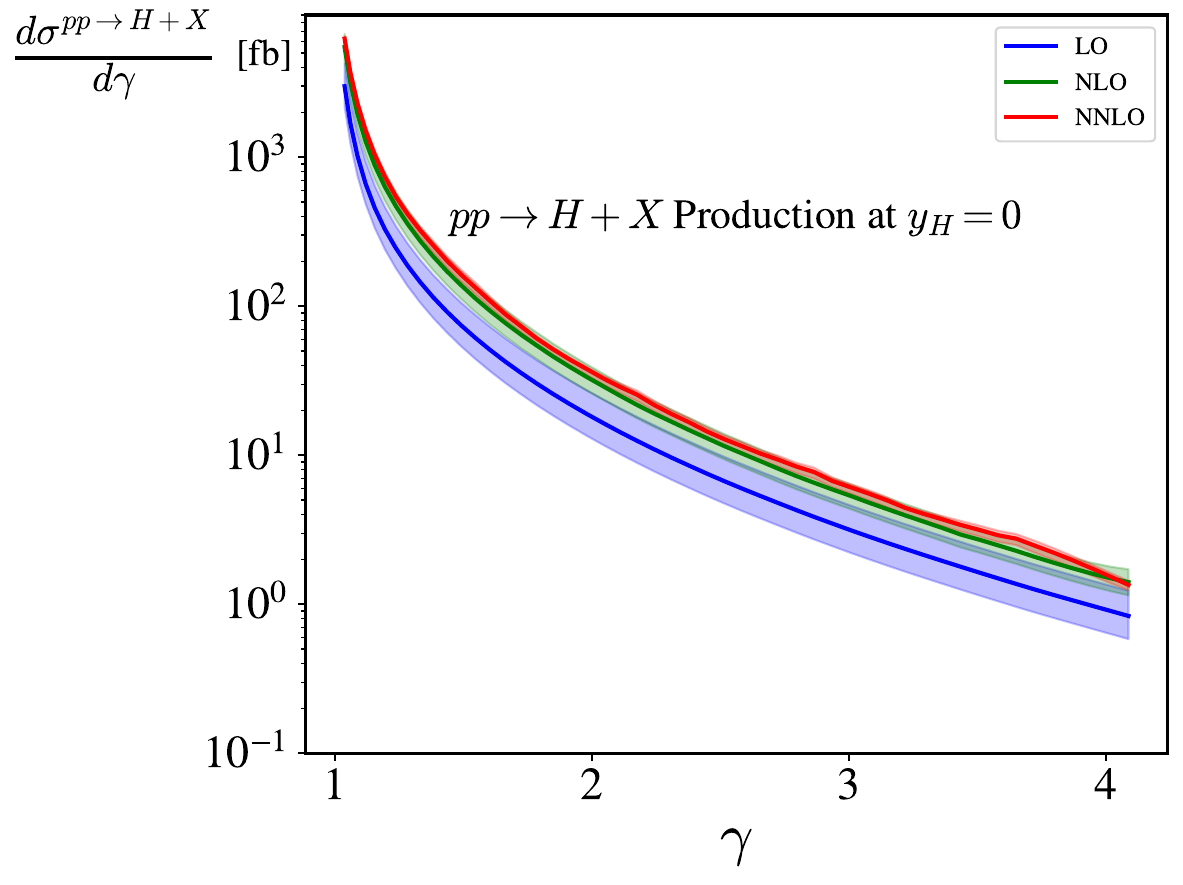}
\end{center}
\caption{Higgs production differential in the boost factor $\gamma = E_H/m_H$ at $y_H=0$ computed using \texttt{NNLOJET}.
The solid line shows the central scale choice, while the shaded band is the envelope of renormalization
and factorization scale variations (see text).}
\label{fig:production}
\end{figure} 
With the narrow-width approximation, the production and decay factorizes as~\eq{proddecay} and the Higgs production cross section
$\sigma^{pp\to H}$ there serves as a proxy for some Higgs production process with a given boost
factor $\gamma = E_H/m_H$.  In this section, as an example, we provide an explicit prediction for the
boost distribution of Higgs production at the LHC with $\sqrt{s}=13$ TeV. We compute the production using the \texttt{HJ} process~\cite{Chen:2014gva,Chen:2016zka} in \texttt{NNLOJET}~\cite{NNLOJET:2025rno}, with the \texttt{PDF4LHC21\_40} PDF set~\cite{PDF4LHCWorkingGroup:2022cjn} accessed via \texttt{LHAPDF6}~\cite{Buckley:2014ana}. In general, our Higgs production is given with some $p_T^{\rm min}<p_T<p_T^{\rm max}$ and $y_H^{\rm min}<y_H<y_H^{\rm max}$. Within such fixed range, we can change our distribution in $p_T$ and $y_H$ to the distribution in a boost factor $\gamma$ via
\begin{align}
\frac{d\sigma^{pp\to H+X}}{d\gamma} = \int_{p_{T,H}^{\rm min}}^{p_{T,H}^{\rm max}} dp_{T,H} \int_{y_H^{\rm min}}^{y_H^{\rm max}} dy_H \frac{d\sigma^{pp\to H+X}}{dp_{T,H} dy_H} \delta\left(\gamma - \frac{\sqrt{p_{T,H}^2 + m_H^2}\,\cosh y_H}{m_H}\right)\,.
\end{align}
Therefore, high $p_{T,H}$ with central rapidity or lower $p_{T,H}$ with high rapidity can both give large boost factor. For simplicity, if we look at the $y_H=0$ slice, the map between $\gamma$ and $p_T$ is simply given by $p_{T,H} = m_H\sqrt{\gamma^2-1}$. Therefore,
\begin{align}
  \frac{d\sigma^{pp\to H+X}}{d\gamma}
  = \frac{d\sigma^{pp\to H+X}}{d p_{T,H}dy_H}\bigg|_{y_H=0}^{ p_{T,H}=m_H\sqrt{\gamma^2-1}}\,
  \frac{m_H\,\gamma}{\sqrt{\gamma^2-1}}\,,
\end{align}
and thus high and low $p_T$ directly corresponds to high and low $\gamma$.

Fig.~\ref{fig:production} shows the LO, NLO, and NNLO predictions for $d\sigma/d\gamma$ at fixed $y_H=0$ slice as a function of $\gamma$. The uncertainties are estimated using $7$-scale variation as in~\cite{Chen:2016zka}: both renormalization scale $\mu_R$ and factorization scale $\mu_F$ are varied around $\mu_0=\frac12\sqrt{m_H^2+p_{T,H}^2}$ by a factor of $2$ and the two extreme variations are excluded.

For the purposes of this work, the production distribution provides an illustrative weighting of Higgs
boost factors.  The steep falloff with $\gamma$ reflects the rapidly decreasing cross section of producing
Higgs bosons at high transverse momentum $p_{T,H}$, while the residual band width indicates that perturbative
uncertainties remain under good control in the $\gamma$ range relevant for the boosted-decay examples in
the main text.




\end{document}